\documentclass[a4paper]{jpconf}

\usepackage[utf8]{inputenc}
\usepackage{graphics,cite,amssymb,epsfig,float}
\usepackage{amsmath}
\usepackage{subfigure}

\usepackage{xcolor} %définir couleur à la main

\def\beq{\begin{equation}}   \def\eeq{\end{equation}}
\newcommand{\Br}{\text{Br}}

\begin{document}
\begin{flushright}
LPT-Orsay-13-106 
\end{flushright}

\title{Enhanced lepton flavour violation in the supersymmetric inverse seesaw}

\author{C Weiland}
\address{Laboratoire de Physique Théorique, CNRS -- UMR 8627, Université Paris Sud 11, F-91405 Orsay Cedex, France}
\ead{cedric.weiland@th.u-psud.fr}

\begin{abstract}
In minimal supersymmetric seesaw models, the contribution to lepton flavour violation from $Z$-penguins is usually negligible. In this study, we consider the supersymmetric inverse seesaw and show that, in this case, the $Z$-penguin contribution dominates in several lepton flavour violating observables due to the low scale of the inverse seesaw mechanism. Among the observables considered, we find that the most constraining one is the $\mu$--$e$ conversion rate which is already restricting the otherwise allowed parameter space of the model. Moreover, in this framework, the $Z$-penguins exhibit a non-decoupling behaviour, which has previously been noticed in lepton flavour violating Higgs decays.
\end{abstract}

While the basic ingredients of neutral lepton flavour violation are close to being fully determined, charged lepton flavour violation (cLFV) remains to be observed despite impressive experimental efforts. In fact, cLVF is allowed in minimal extensions of the Standard Model but the associated observables are highly suppressed due to a Glashow–Iliopoulos–Maiani (GIM) mechanism taking place in the lepton sector. However, when these extensions are embedded in a larger framework like supersymmetry (SUSY), large contributions to cLFV are expected. The supersymmetric inverse seesaw is a very attractive framework to consider since all the new Physics it introduces does not exceed the TeV scale. As a consequence, it could be tested at high-energy through the direct production of the new heavy particles but also at the high-intensity frontier. Indeed, in the inverse seesaw \cite{Mohapatra:1986bd}, the size of the effective operator at the origin of the neutrino masses is not correlated to the size of the one generating cLFV. This suggests that there could be a large enhancement to lepton flavour violating observables, which has recently been discussed in several studies \cite{Deppisch:2004fa,Deppisch:2005zm,Dev:2009aw,Hirsch:2009ra,Dias:2012xp,Hirsch:2012kv,Awasthi:2013we}. As recently pointed out, Higgs-mediated observables can be significantly enhanced by the presence of relatively light right-handed neutrinos \cite{Abada:2011hm}. Contrary to the Minimal Supersymmetric Standard Model (MSSM) where $Z$-penguin contributions are suppressed due to several cancellations, in the supersymmetric inverse seesaw, $Z$-mediated contributions are enhanced and even dominate the ones arising from Higgs-penguin. Indeed, in most of the cases, the $Z$-penguins provide the dominant contribution and can also exhibit a surprising non-decoupling behaviour.

After defining the model, we will discuss the enhancement of the $Z$-mediated contribution, giving an approximate formula for the $Z$--$\ell_i$--$\ell_j$ vertex. We will then present our numerical results and conclusions.

\section{The supersymmetric inverse seesaw}

The supersymmetric inverse seesaw consists of the MSSM extended by three pairs of gauge singlet superfields, $\widehat{\nu}^c_i$ and
$\widehat{X}_i$ ($i=1,2,3$), with opposite lepton number, $-1$ and $+1$, respectively. The superpotential of this model is given by

\begin{align}
{\mathcal W}&= \varepsilon_{ab} \left[
Y^{ij}_d \widehat{D}_i \widehat{Q}_j^b  \widehat{H}_d^a
              +Y^{ij}_{u}  \widehat{U}_i \widehat{Q}_j^a \widehat{H}_u^b 
              + Y^{ij}_e \widehat{E}_i \widehat{L}_j^b  \widehat{H}_d^a \right. \nonumber \\
              &+\left. Y^{ij}_\nu 
\widehat{\nu}^c_i \widehat{L}^a_j \widehat{H}_u^b - \mu \widehat{H}_d^a \widehat{H}_u^b \right] 
+M_{R_{ij}}\widehat{\nu}^c_i\widehat{X}_j+
\frac{1}{2}\mu_{X_{ij}}\widehat{X}_i\widehat{X}_j  ~,
\label{eq:SuperPot}
\end{align}

\noindent where $i,j = 1,2,3$ are generation indices.  In the above,
$\widehat H_d$ and $\widehat H_u$ are the down- and up-type Higgs
superfields, $\widehat L_i$ denotes the SU(2) doublet lepton
superfields. The only term that violates lepton number conservation is the Majorana mass term $\mu_{X_{ij}}\widehat{X}_i\widehat{X}_j $. Thus, the limit $\mu_{X_{ij}}\rightarrow 0$ increases the symmetry of the model, making the smallness of $\mu_{X_{ij}}$ natural. The soft SUSY breaking Lagrangian can be found in \cite{Abada:2012cq}. Assuming a flavour-blind mechanism of supersymmetry breaking by considering universal boundary conditions at the grand unification (GUT) scale, like in case of the constrained MSSM (cMSSM),  the supersymmetric parameters are

\beq m_\phi = m_0\,, M_\text{gaugino}= M_{1/2}\,,
A_{i}= A_0 \, \mathbb{I}\,,B_{\mu_X}=B_{M_R}= B_0 \, \mathbb{I}\,.  \eeq

\noindent In this framework, the $9\times9$ neutrino mass matrix $\mathcal{M}$ in the basis $\{\nu,{\nu^c},X\}$ is given by 

\begin{eqnarray}
{\cal M}&=&\left(
\begin{array}{ccc}
0 & m^{T}_D & 0 \\
m_D & 0 & M_R \\
0 & M^{T}_R & \mu_X \\
\end{array}\right) \ ,
\label{nmssm-matrix}
\end{eqnarray}

\noindent where $m_D= \frac{1}{\sqrt 2} Y_\nu v_u$ and $M_R$, $\mu_X$ are  
$3\times3$ matrices in family space.
Assuming $m_D,\mu_X \ll M_R$,  the diagonalization leads to an
effective Majorana mass matrix for the light
neutrinos~\cite{GonzalezGarcia:1988rw},

\begin{equation}
\label{eqn:nu}
    m_\nu = {m_D^T M_R^{T}}^{-1} \mu_X M_R^{-1} m_D
          = \frac{v_u^2}{2} Y^T_\nu (M^T_R)^{-1} \mu_X M_R^{-1} Y_\nu\,.
\end{equation}

As can be seen from the above equation, the smallness of the light neutrino masses can be explained by the smallness of $\mu_X$. This decorrelates the effective operator generating neutrino masses from the one generating cLFV since the latter does not depend on $\mu_X$. Moreover having $\mu_X$ at the eV-scale allows the Yukawa couplings to have natural values ($Y_\nu \sim \mathcal{O}(1)$) and, at the same time, right-handed neutrinos to have masses at the TeV. Having light right-handed neutrinos and large Yukawa couplings leads, in turn, to large contributions to cLFV observables. By defining 

\begin{equation} 
\label{M-def}
M^{-1} = (M^T_R)^{-1} \ \mu_X \ M_R^{-1}\,,
\end{equation}

\noindent the light neutrino mass matrix can be cast in a form that strongly resembles the one for the type-I seesaw

\begin{equation} \label{M-def2}
 m_\nu = \frac{v_u^2}{2} Y^T_\nu M^{-1}  Y_\nu\,.
\end{equation}

Without loss of generality, we work in a basis where $M_R$ is diagonal at the SUSY scale. Moreover, in the numerical analysis, we will work in the limit where $\mu_X$ is diagonal. This is a conservative assumption, since, in that case, the experimental limits will constrain the size of the Yukawa couplings, which in turn will limit the branching ratios and decay rates of cLFV observables. The constraints from the neutrino data that we considered in this study can be found in \cite{Abada:2012cq}.

\section{$Z$-mediated charged lepton flavour violation}

Many studies have been devoted to cLFV in the MSSM and many of its extensions. It was found that the $\gamma$-penguin gives the dominant contribution to 3-body cLFV decays \cite{Hisano:1995cp,Arganda:2005ji}, the only exception being the large $\tan \beta$ (and low $m_A$) regime where large Higgs contributions are expected \cite{Babu:2002et}. The same observation was recently made in the supersymmetric inverse seesaw\cite{Abada:2011hm}. When the mass scaling of the different contributions is considered, the $\gamma$-penguin contribution to $\ell_i \rightarrow 3 \ell_j$ from the chargino-sneutrino loop is given by

\begin{equation} \label{Achar}
A_a^{(c)L,R} = \frac{1}{16 \pi^2 m_{\tilde{\nu}}^2} {\cal O}_{A_a}^{L,R}s(x^2)\, ,  
\end{equation}
whereas the $Z$-contribution reads
\begin{equation} \label{Fchar}
F_{X} = \frac{1}{16 \pi^2 g^2 \sin^2 \theta_W m^2_{Z}}{\cal O}_{F_X}^{L,R}t(x^2)\, ,
\end{equation}

\noindent with $X=\left\{LL,LR,RL,RR\right\}$.  In the above, ${\cal
  O}_{y}^{L,R}$ denotes combinations of rotation matrices and coupling
constants and $s(x^2)$ and $t(x^2)$ represent the Passarino-Veltman
loop functions which depend on $x^2 \equiv
m_{\tilde{\chi}^-}^2/m_{\tilde{\nu}}^2$ (see
\cite{Arganda:2005ji}). Since $m_{Z}^{2} \ll m_{\tilde{\nu}}^2$, the $Z$-mediated contribution would dominate over the one coming from $\gamma$-penguins. However, in the MSSM, this is prevented by a subtle cancellation between different diagrams contributing to the leading $Z$-mediated contribution~\cite{Hirsch:2012ax}. It can be understood by focusing on the dominant contribution to $\ell_i \to 3 \ell_j$ which is given by diagrams where the external leptons are left-handed (the other cases are suppressed by the Yukawa couplings of the charged
leptons). It is given in \cite{Abada:2012cq} by the form factor

\begin{equation}
\label{FLLfirst}
F_{LL} = \frac{F_L Z_L^{(l)}}{g^2 \sin^2 \theta_W m_Z^2}\ ,
\end{equation}

\noindent where $Z_L^{(l)} = - \frac{g}{c_W}(-\frac{1}{2} + s_W^2)$ is the
$Z-\ell_i-\ell_j$ tree-level coupling ($i=j$) and $F_L$ 
is the $Z-\ell_i-\ell_j$ 1-loop effective
vertex, with $i \ne j$. In the MSSM, the sneutrino mass matrix is diagonalized by $Z_V$, a $3 \times 3$ unitary matrix. But when chargino mixing is neglected, the masses cancel out in the combination of loop functions from different diagrams and one ends up with $F_L^{0}\propto(Z^\dagger_{V} Z_{V})_{ij}$ which vanishes for $i \ne j$ due to the unitarity of the $Z_V$ matrix\cite{Hirsch:2012ax}. However, this cancellation can be spoiled if new interactions are introduced like it is the case with the inverse seesaw mechanism+. In particular, the $\tilde{H}^\pm-\tilde{\nu}_R-\ell_L$ coupling gives rise to new diagrams containing a higgsino-sneutrino loop which spoils the cancellation since they give $F_L^{0} \propto \sum_i Z_{V}^{ik} Z_{V}^{ij*} Y_\nu^{ik*} Y_\nu^{ij}$. It is worth noticing that this does not happen in the MSSM extended by a high-scale type-I seesaw since the very heavy right-handed sneutrinos suppress the higgsino-sneutrino loop. In the supersymmetric inverse seesaw, the large neutrino Yukawa couplings and the TeV-scale right-handed sneutrinos make this contribution naturally large.

The complete derivation of the analytical expressions for $F_L^{0}$ can be found in \cite{Abada:2012cq} and only an abridged version will be presented here. The chargino-sneutrino loops can be decomposed as

\begin{equation}
\label{FL0pi}
F^{(0)}_L = - \frac{1}{16 \pi^2} \left( F^{(0)}_{L,\text{wino}} + F^{(0)}_{L,\text{higgsino}} \right) \,.
\end{equation}

\noindent Considering that left-right mixing is negligible in the sneutrino sector when the neutrino Yukawa couplings are relatively small (which will be verified by our numerical study), the two contributions reduce to

\begin{eqnarray}
\label{FL0wino}
F^{(0)}_{L,\text{wino}} &=& \frac{g^2}{8} \delta_{ij} (3 g c_W + g' s_W) = \frac{g^3}{8 c_W} \delta_{ij} (1 + 2 c_W^2)\,, \\
F^{(0)}_{L,\text{higgsino}} &=& \frac{g}{8 c_W} \left( Y_\nu^\dagger Y_\nu \right)_{ij} \left( c_W^2 - \frac{1}{2} \right)\,.
\end{eqnarray} 

\noindent It can be noted that the wino contribution vanishes for $i\ne j$, which corresponds to the cancellation discussed earlier, while the higgsino contribution, which is absent from the MSSM, does not vanish, explaining the different behaviours between the two models. We have verified that the expressions above reproduce the numerical results based on complete, non-approximate expressions. We have also verified that the higher-order terms in chargino mixing are numerically negligible. It is worth mentioning that the $F_L^{(0)}$ vertex does not depend on supersymmetric masses. As a consequence, the $Z$-penguin exhibits a non-decoupling behaviour and a large SUSY scale does not suppress its contribution to cLFV processes.

Finally, although the previous discussion has been focused on $\ell_i \to 3 \ell_j$ processes,
the same enhancement in the $Z-\ell_i-\ell_j$ effective vertex
also affects other observables. 
This is the case for $\mu-e$ conversion in nuclei \cite{Arganda:2007jw} 
and the case of $\tau \to P^0 \ell_i$, where $P^0$ is a pseudoscalar meson
\cite{Arganda:2008jj}.

\section{Numerical results}

\begin{table}[tb!]
\centering
\begin{tabular}{|c|c|c|}
\hline
cLFV Process & Present Bound & Future Sensitivity  \\
\hline
$\mu \to e \gamma$ & $2.4 \times 10^{-12}$ \cite{Adam:2011ch} & $6\times10^{-14}$ \cite{Baldini:2013ke}  \\
$\tau \to e \gamma$ & $3.3 \times 10^{-8}$ \cite{Aubert:2009ag}& $3.0 \times 10^{-9}$ \cite{OLeary:2010af}\\
$\tau \to \mu \gamma$ & $4.4 \times 10^{-8}$ \cite{Aubert:2009ag}& $2.4 \times 10^{-9}$ \cite{OLeary:2010af} \\
$\mu \to 3 e$ & $1.0 \times 10^{-12}$\cite{Bellgardt:1987du} & $\mathcal{O}(10^{-16})$ \cite{Blondel:2013ia}\\
$\tau \to 3 e$ & $2.7\times10^{-8}$\cite{Hayasaka:2010np} & $2.3 \times 10^{-10}$ \cite{OLeary:2010af}  \\
$\tau \to 3 \mu$ & $2.1\times10^{-8}$\cite{Hayasaka:2010np} & $8.2 \times 10^{-10}$ \cite{OLeary:2010af}  \\
$\mu-e$ , Au & $7.0 \times 10^{-13}$ \cite{Bertl:2006up} & $ $  \\
$\mu-e$ , Ti & $4.3 \times 10^{-12}$ \cite{Dohmen:1993mp} & $\mathcal{O}(10^{-18})$ \cite{Barlow:2011zza} \\
$\tau \to e \eta$ & $4.4\times 10^{-8}$\cite{collaboration:2010ipa} & $\mathcal{O}(10^{-10})$ \cite{OLeary:2010af}\\
$\tau \to \mu \eta$ & $2.3\times 10^{-8}$\cite{collaboration:2010ipa} & $\mathcal{O}(10^{-10})$ \cite{OLeary:2010af}\\
\hline
\end{tabular}
\caption{Current experimental
  bounds and future sensitivities for different cLFV observables.}
\label{tab:sensi}
\end{table}

In this study, we focused on leptonic observables which can be divided in three broad classes: radiative decays, like $\mu \rightarrow e \gamma$, 3-body decays, like $\tau \rightarrow 3 \mu$, and neutrinoless conversion in muonic atoms, like $\mu,\mathrm{Ti}\rightarrow e,\mathrm{Ti}$. These observables are actively searched for at many current and future experiments. The list of processes we considered together with the current upper limit and expected future sensitivity can be found in table \ref{tab:sensi}. We also addressed high-energy observables such as flavour-violating neutralino decays $\Br(\tilde{\chi}_2^0 \to\tilde{\chi}_1^0 \ell_i \ell_j)$ and slepton mass splittings $\Delta m_{\tilde{\ell}}$ ($m_{\tilde{\ell_i}}-m_{\tilde{\ell_j}}$). However, we found that the small neutrino Yukawa couplings needed to comply with experimental limits on $\mu,\mathrm{N} \rightarrow e,\mathrm{N}$ conversion (as discussed below) gives low  rates for $\tilde{\chi}_2^0 \to \tilde{\chi}_1^0 \ell_i \ell_j$ and small mass splittings. We have also verified that cLFV decays of the $Z$ and $h^0$ bosons are not significantly enhanced.

\begin{figure}
\centering
\subfigure[$^{48}_{22}\text{Ti}$]{
\includegraphics[width=0.45\linewidth]{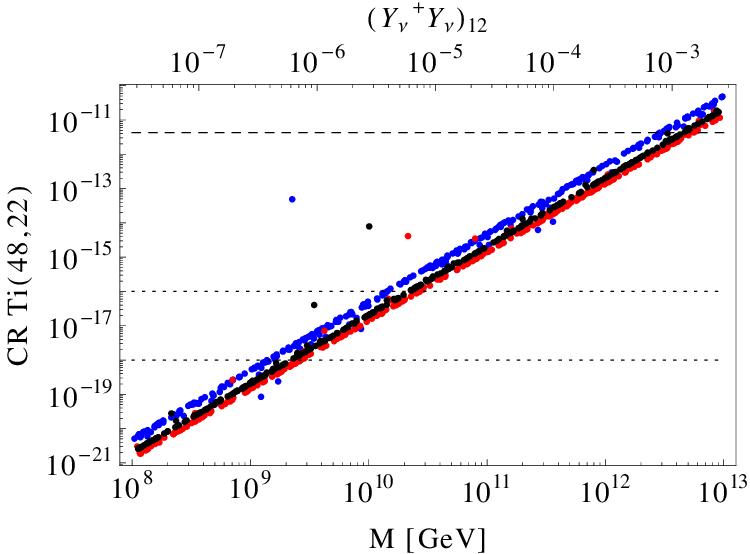}
\label{CRTi-scatter}
}
\subfigure[$\mu \to e \gamma$]{
\includegraphics[width=0.5\linewidth]{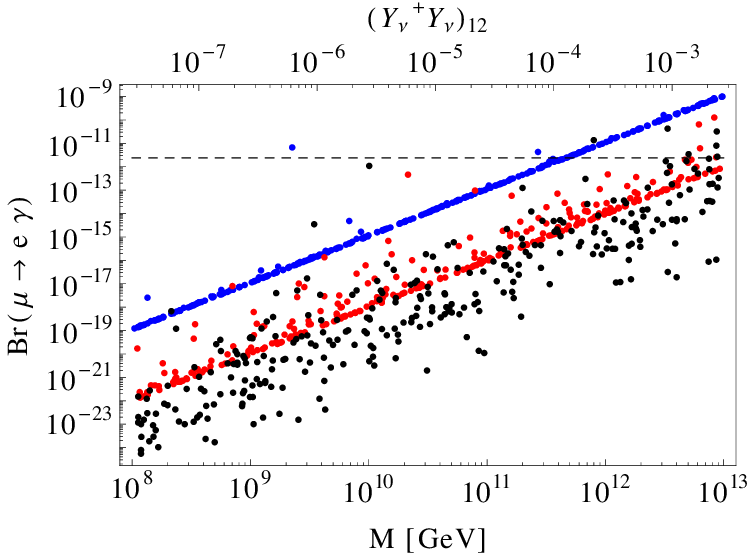}
\label{muegamma-scatter}
}
\caption{$\mu-e$ conversion rates in $^{48}_{22}\text{Ti}$ (left) and $\Br(\mu \to e \gamma)$ (right), as a function of $M$ and $(Y_\nu^\dagger Y_\nu)_{12}$ for different $\hat{M}_R$ values: $\hat{M}_R = 100$ GeV (blue), $\hat{M}_R = 1$ TeV (red) and $\hat{M}_R = 10$ TeV (black). We set $A_0 = -300$ GeV, $\tan \beta = 10$, $\text{sign}(\mu) = +$ and $B_0 = 0$,  and we randomly  vary $m_0$ and $M_{1/2}$  in the range [$0,3$ TeV]. The horizontal dashed lines represent the current experimental bounds and the dotted ones represent the expected future sensitivities.}
\label{CRfig}
\end{figure}

Our numerical results have been obtained with a {\tt SPheno} \cite{Porod:2003um,Porod:2011nf} code generated with the Mathematica package {\tt SARAH} \cite{Staub:2011dp,Staub:2010jh,Staub:2009bi}. The computation of the cLFV observables is based on the results presented in~\cite{Arganda:2005ji,Arganda:2007jw,Arganda:2008jj}, that we extended to the inverse seesaw case. As such, it is not limited to the $Z$-penguins but also includes contributions from Higgs-penguins, $\gamma$-penguins and supersymmetric boxes. The analytical expressions for $\mu-e$ conversion in nuclei, $\Br(\ell_i \to 3 \ell_j)$ and $\Br(\tau \to P^0 \ell_i)$ in the case of $Z$-penguin dominance are given in \cite{Abada:2012cq}. The supersymmetric parameters of the model are $m_0, M_{1/2}, A_0, \tan\beta, \text{sign}(\mu)$ and $B_0$. Since they are of little relevance for $Z$-mediated observables, we fixed $\text{sign}(\mu)= +$  and $B_0=0$ in the numerical analysis. As can be seen in figure \ref{CRTi-scatter}, when an observable receives its dominant contribution from $Z$-penguins, it has a very small dependence on the cMSSM parameters making our results quite general. However, this is not true when another contribution is dominant like in $\mathrm{Br}(\mu \rightarrow e \gamma)$, see figure \ref{muegamma-scatter} for example, in accordance with the literature \cite{Deppisch:2004fa,Hirsch:2009ra}.   The results presented here are obtained with degenerate singlets $M_R = \text{diag}(\hat{M}_R,\hat{M}_R,\hat{M}_R)$ and we have verified that processes dominated by $Z$-penguins are only marginally affected by non-degenerate singlets. For example, the relative change in the conversion rate for $\mu,\mathrm{Au}\rightarrow e,\mathrm{Au}$ is always below $2\%$. As a consequence, the results discussed here hold true for both degenerate and non-degenerate right-handed (s)neutrinos. Since $\mu_X$ does not explicitely appear in analytical expressions for cLFV processes, the relevant parameters are $M \sim \frac{M_R^2}{\mu_X}$ (as defined in Eq. \eqref{M-def}), which controls the size of the Yukawa couplings, and $M_R$ which controls the size of the non-SUSY contributions.

\begin{figure}
\centering
\subfigure[$\, \mu \to 3 e$]{
\includegraphics[width=0.45\linewidth]{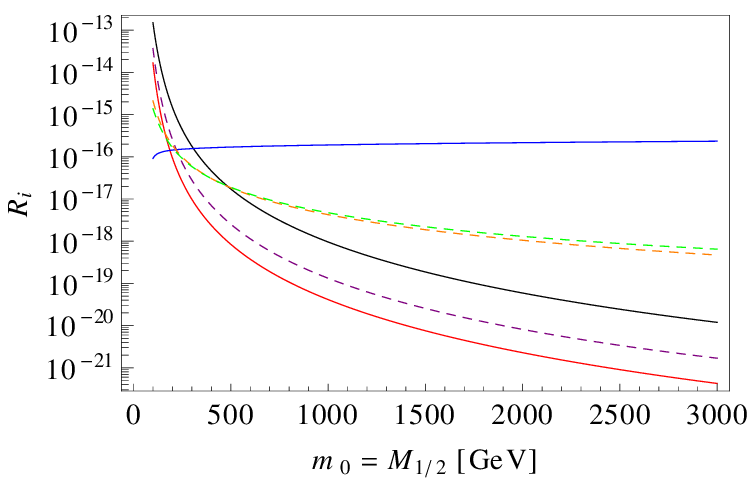}
\label{contMEEE}
}
\subfigure[$\, \mu-e$ conversion in $^{197}_{79}\text{Au}$]{
\includegraphics[width=0.45\linewidth]{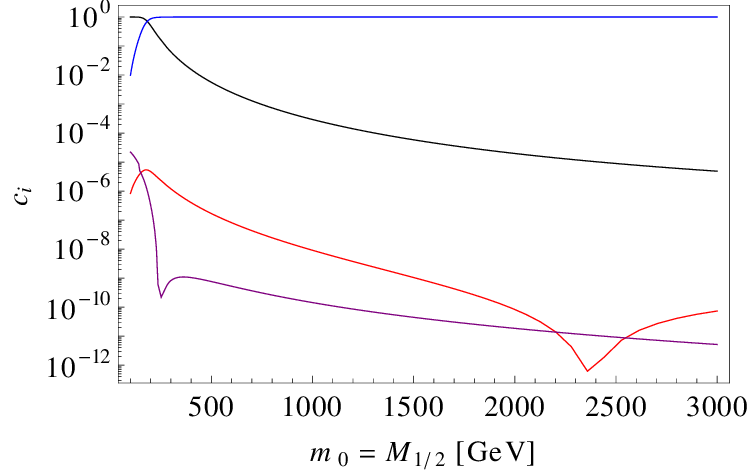}
\label{contMUE}
}
\caption{Absolute contributions to $\Br(\mu \to 3 e)$ (left-hand side) and relative contributions to $\mu-e$ conversion in $^{197}_{79}\text{Au}$ (right-hand side) as a function of $m_0 = M_{1/2}$ for a degenerate singlet spectrum with $\hat{M}_R = 10$ TeV and $M = 10^{11}$ GeV. The rest of the cMSSM parameters are set to  $A_0 = -300$ GeV, $B_0 = 0$, $\tan \beta = 10$ and $\text{sign}(\mu)=+$. On the left-hand side, solid lines represent individual contributions, $\gamma$ (black), $Z$ (blue) and $h$ (red) whereas the dashed lines represent interference terms, $\gamma-Z$ (green), $\gamma-h$ (purple) and $Z-h$ (orange). Note that in this case $h$ includes both Higgs and box contributions. On the right-hand side interference terms are not shown to make the results clearer. The individual contributions are $\gamma$ (black), $Z$ (blue) and $h$ (red) and boxes (purple).}
\label{fig:contributions}
\end{figure}

In figure \ref{contMEEE}, we display the absolute contributions to $\mathrm{Br}(\mu \rightarrow 3 e)$ in order to illustrate the non-decoupling behaviour of the $Z$-penguin as previously discussed. We define the relative contribution to cLFV observables as

\begin{equation} \label{def-ci}
c_i = \frac{|\text{R}_i|}{\sqrt{\sum_k \text{R}_k^2}},
\end{equation}

\noindent where $\text{R}_i$ is the rate (branching ratio in case of $\mu \to 3 e$ and conversion rate in case of $\mu-e$ conversion in Au) that would be obtained, should the $i$-contribution to the process be the only one. It can be seen from figure \ref{contMUE} that $\mathrm{Br}(\mu \rightarrow 3 e)$ is clearly dominated by the $Z$-boson contribution in most of the parameter space, the only exception being for very low $m_0=M_{1/2}$ where the $\gamma$-penguins dominate. This can easily be understood by recalling the scaling of the operators for the different contributions in equations \eqref{Achar} and \eqref{Fchar}. The same behaviour was observed in other observables as described in \cite{Abada:2012cq} and allows to conclude that,  if present, the $Z$-mediated contribution will dominate cLFV observables in the supersymmetric inverse seesaw, except for a very low SUSY scale.

When the $Z$-penguin gives the dominant contribution, there is a strong correlation between $M$ and the processes rate. This can clearly be seen in figure \ref{CRTi-scatter} for the $\mu - e$ conversion rate. From the non-observation of cLFV, an upper bound on $M$, or equivalently on the size of $(Y_\nu^\dagger Y_\nu)_{ij}$, can be derived. Among the many observables involving $\mu - e$ transitions, the most constraining one is neutrinoless conversion in muonic atoms and the approximate limits it gives are summarised in table \ref{tab:muebounds} for a conservative choice of $\hat{M}_R \gtrsim 1\, \mathrm{TeV}$. Smaller values for $\hat{M}_R$ will only result in stronger limits, especially when non-SUSY contributions become dominant, making $\mu \rightarrow e \gamma$ the most constraining observable. It should also be noted that future experiments like COMET, PRISM/PRIME and Mu2e have the potential to greatly improve these limits.

\begin{table}[hbt]
\centering
\begin{tabular}{c c c}
\hline
 & $M$ [GeV] & $(Y_\nu^\dagger Y_\nu)_{12}$ \\
\hline
$\text{CR}_{\text{Au}}$(current) & $10^{12}$ & $2.7  \times 10^{-4}$ \\
$\text{CR}_{\text{Ti}}$(current) & $4  \times 10^{12}$ & $10^{-3}$ \\
$\text{CR}_{\text{Ti}}$(future: $10^{-16}$) & $2  \times10^{10}$ & $5.5 \times10^{-6}$ \\
$\text{CR}_{\text{Ti}}$(future: $10^{-18}$) & $2  \times 10^{9}$ & $5.5 \times10^{-7}$ \\
\hline
\end{tabular} 
\caption{Approximate limits on $M$ and $(Y_\nu^\dagger Y_\nu)_{12}$ from the non-observation of $\mu-e$ conversion in nuclei. This table includes current experimental bounds and future expected sensitivities~\cite{Bertl:2006up, Dohmen:1993mp, Barlow:2011zza, Glenzinski:2010zz,Cui:2009zz,Carey:2008zz}.}
\label{tab:muebounds}
\end{table}

Other cLFV transitions also exhibit a clear enhancement due to the $Z$-penguin dominance. In particular, the branching ratios of different $\ell_i \to 3 \ell_j$ channels are related by

\begin{equation} \label{ratioBRs}
\frac{\Br(\ell_i \to 3 \ell_j)}{\Br(\ell_m \to 3 \ell_n)} = \frac{(Y_\nu^\dagger Y_\nu)_{ij}^2}{(Y_\nu^\dagger Y_\nu)_{mn}^2} \frac{m_{\ell_i}^5 \tau_i}{m_{\ell_m}^5 \tau_m}\,.
\end{equation}

\noindent From this equation, we can expect that, considering similar combination $(Y_\nu^\dagger Y_\nu)_{ij}$, all the branching ratios for $\ell_i \to 3 \ell_j$ processes will lie with $1-2$ orders of magnitude. However, the experimental upper limits on these observables are more relaxed for $\tau - e$ and $\tau - \mu$ cLFV transitions than for $\mu - e$ cLFV transition. This implies that 3-body $\tau$ cLFV decays have little possibility to be observed in the near future, except if there is a strong cancellation in $Y_\nu^\dagger Y_\nu$ which would suppress $\mu - e$ cLFV processes. The same conclusion also holds for semileptonic $\tau$ decays. Finally, it is worth mentioning that in $Z$-penguin dominated scenarios, there is a clear correlation between the rates for $\mu \to 3 e$ and $\mu-e$ conversion in nuclei. Numerically, we found $\text{CR}(\mu-e,\text{Ti})/\Br(\mu \to 3e) \sim 15$.

$Z$-dominated observables are quite insensitive to the cMSSM parameters, implying that these results should hold when the recent mass constraints on the Higgs boson are imposed on the model. We have checked this using a benchmark point associated to a Higgs mass of $m_{h^0}=126.5\,\mathrm{GeV}$ and the corresponding predictions for the different observables are given in \cite{Abada:2012cq}. Moreover, since the neutrino Yukawa couplings are constrained to be below $10^{-2}$, we do not expect the singlet sector to sensibly modify the Higgs mass and renormalisation group equations (RGEs).\\

To conclude, the supersymmetric inverse seesaw is a very attractive extension of the MSSM since it generates neutrino masses while keeping all the new Physics below the TeV scale with naturally large Yukawa couplings. In this work, we considered its predictions for different lepton flavour violating observables mediated by the exchange of a $Z$-boson and found that the $Z$-mediated contribution dominates in most of the parameter space, providing an important enhancement to different observables. As a consequence, the most constraining observable is now neutrinoless $\mu - e$ conversion in nuclei, in contrast with the usual behaviour of the MSSM extended by a high-scale type-I seesaw. This allows us to put constraints on the size of the Yukawa couplings $(Y_\nu^\dagger Y_\nu)_{12} \gtrsim 2.7 \times 10^{-4}$, clearly excluding $Y_\nu\sim \mathcal{O}(1)$ as usually considered.

\section*{Acknowledgements}

This work has been supported by the ANR project CPV-LFV-LHC NT09-508531 and the European Union FP7 ITN INVISIBLES (Marie Curie Actions, PITN-GA-2011-289442).

\vspace*{0.5cm}

\bibliographystyle{iopart-num}
\bibliography{Discrete}

\end{document}